\documentstyle[preprint,tighten,aps,epsfig]{revtex}
\newcommand{\gz}{$e^+e^- \rightarrow \gamma Z$}
\newcommand{\zz}{$e^+e^- \rightarrow Z Z$}
\newcommand{\bq}{\begin{equation}}
\newcommand{\eq}{\end{equation}}

\newcommand{\dslash}{\hbox{$\partial$\kern-0.5em\raise0.3ex\hbox{/}}}
\newcommand{\Dslash}{\hbox{$D$\kern-0.5em\raise0.3ex\hbox{/}}}
\newcommand{\eqn}[1]{Eq.~(\ref{#1})}
\def\msb{$\overline{\rm{MS}}$ }
\def\mt{m_t}
\def\mz{m_Z}
\def\m{\mu}

\def\4pi{(4 \pi)}
\def\gf{G_F}
\def\logm{\log\left(\frac{m_t}{m_Z}\right)}

\def\ZZ{$e^+e^- \rightarrow Z Z$ }
\def\gZ{$e^+e^- \rightarrow \gamma Z$ }
\begin{document}

\preprint{
\begin{tabular}{r} FTUV/99$-$15 \\ IFIC/99$-$16
\end{tabular}
}

\title{An effective field theory approach to the electroweak corrections
       at LEP energies}
\author{A.\ A.\ Akhundov\footnote{On sabbatical leave from {\em Institute
of Physics, Azerbaijan Academy of Sciences, AZ-370143 Baku, Azerbaijan}},
        J.\ Bernab\'eu, D.\ G\'omez Dumm and A.\ Santamaria}
\address{\hfill \\ Departament de F\'{\i}sica Te\`orica, IFIC, 
CSIC -- Universitat de Val\`encia \\
Dr.\ Moliner 50, E-46100 Burjassot (Val\`encia), Spain}

\maketitle

\thispagestyle{empty}
 
\begin{abstract}
In the framework of the effective field theory (EFT) we discuss the
electroweak (EW) corrections at LEP energies. We obtain the
effective Lagrangian in the large $m_t$ limit, and reproduce analytically
the dominant EW corrections to the LEP2 processes
$e^+e^- \rightarrow \gamma Z$
and $e^+e^- \rightarrow Z Z$. To include effects of finite top--quark and
Higgs masses, we use the effective Lagrangian at tree level
and fit LEP1/SLD observables with four arbitrary parameters, plus
$\alpha_s(m_Z)$. The EFT approach works remarkably well. Using the
effective couplings determined from the fit, and tree--level EFT
formulae, we predict the cross sections
for $e^+e^- \rightarrow Z Z$, $\gamma Z$ at a level better than~1\%.
\end{abstract}

\hfill


\renewcommand{\baselinestretch}{1}

\section{Introduction}
\label{sec:intro}

The radiative corrections play a very important role in the analysis
of the Standard Model (SM) predictions~\cite{PDG,langacker}. 
The knowledge of radiative corrections up to definite order for
different processes is necessary to perform accurate tests of the
SM, allowing to probe the quantum structure of the theory and
also to search for possible effects of new
physics~\cite{hollik}. The precision achieved in these tests has
been significantly increased in the last years, in view of the
experimental information provided by the $e^+e^-$ colliders LEP and SLC and
the $\bar pp$ collider Tevatron~\cite{karlen}, and the theoretical
computation of the SM predictions including radiative corrections beyond
the level of one loop~\cite{hollik,harlander,stuart}. In fact, the calculation
of higher--order radiative corrections has reached an extremely complicated
level and heavily relies now on computer facilities~\cite{steinhauser}. For
the main processes measured at LEP1, the final results are presented through
computer programs~\cite{EWWG}.

The main goal of this paper is to show how the effective field
theory (EFT)~\cite{georgi,peccei,santi,santam} can help in the estimation 
of the electroweak (EW) corrections at LEP2 energies by using the precise
measurements of LEP1 and SLD. Indeed, the standard approach to EW radiative
corrections in the SM requires firstly the evaluation of those corrections
for LEP1/SLD observables, in order to extract the relevant SM parameters from
the available experimental data.
It is seen that the extracted values depend strongly on the top--quark mass,
$\mt$, and (to a lesser extent) on the Higgs mass, $m_H$. Then, in terms of
these parameters,
one can calculate the radiative corrections to LEP2 observables. As expected,
the results also show a strong dependence on $\mt$ and $m_H$. However,
this is cancelled almost completely by the $\mt$ and $m_H$
dependences of the input parameters extracted from LEP1/SLD. This is not
surprising, since for both LEP1 and LEP2 energies top quarks and Higgs
bosons are always virtual. It is then conceivable that a description
in terms of an effective theory without explicit top quarks or Higgs bosons
is good enough for both LEP1 and LEP2. All top quark and Higgs--boson mass
dependences will be absorbed in the effective Lagrangian parameters, which
can be determined at LEP1/SLD and then used to make predictions for LEP2 that
will be trivially independent on $\mt$ and $m_H$. We will show
that this program can be carried out basically at tree level, achieving
precisions for LEP2 predictions at the \% level, which should be enough
for most purposes.

We will focus on the neutral gauge boson production at LEP2,
\bq
e^+ e^- \rightarrow \gamma Z
\label{eq:pgz}
\eq
and 
\bq
e^+ e^- \rightarrow Z Z \;.
\label {eq:pzz}
\eq
The process (\ref{eq:pgz}), so--called ``Z radiative return'', has already
been observed at LEP2 with a hard photon and an on--shell $Z$ in the
final state~\cite{L3} and it is expected that about 10000 $\gamma Z$
events will be collected until the end of LEP2. This process is the main
source of the well--known Initial State Radiation in $e^+e^-$ 
collisions~\cite{LEP2}. The process (\ref{eq:pzz}) is similar to
(\ref{eq:pgz}), with a lower cross section~\cite{zzl3}. Both processes 
(\ref{eq:pgz})
and (\ref{eq:pzz}) are interesting for seeking for possible nonstandard
effects in LEP2, such as the presence of anomalous three--gauge boson
couplings~\cite{abraham}.

The complete one--loop EW corrections in the SM for (\ref{eq:pgz}) and
(\ref{eq:pzz}) have been calculated some years ago by B\"ohm and
Sack~\cite{bohm} and by Denner and Sack~\cite{denner} respectively,
using an on--shell renormalization scheme~\cite{hubert} in
the 't Hooft--Feynman gauge. For both processes (\ref{eq:pgz}) and
(\ref{eq:pzz}), the differential cross section can be written as
\bq
\frac{d\sigma}{d\Omega} =
\left(\frac{d\sigma}{d\Omega}\right)_{0}(1 +\delta^{\rm{QED}}
+\delta^{\rm EW}) \,,
\label{delta}
\eq
where $(d\sigma/d\Omega)_{0}$ is the corresponding cross section in the
Born approximation.
For convenience, the one--loop corrections have been split in two:
$\delta^{\rm{QED}}$ contains the ``pure'' QED ---or photonic---
contributions,
while $\delta^{\rm EW}$ includes the remaining, 
non--QED EW corrections.

The pure QED correction $\delta^{\rm{QED}}$ for neutral gauge boson production
is gauge invariant and depends on the experimental conditions. To get an
analytical result, it is possible to use the soft--photon approximation for
the Bremsstrahlung of an additional photon with energy
$\omega\alt\Delta E$, being $\Delta E$ some energy cutoff. In this way
$\delta^{\rm{QED}}$ has been calculated for the processes
(\ref{eq:pgz}) and (\ref{eq:pzz}) in~\cite{berends} and~\cite{denner}
respectively. It is seen that, with a cutoff $\Delta E=0.1\sqrt{s}$,
$\delta^{\rm{QED}}$ can reach up to $\sim -10$\% for process (\ref{eq:pgz})
and $\sim -17$\% for process (\ref{eq:pzz}) at LEP2 energies. In this paper
we will concentrate in the analysis of the non--QED corrections,
$\delta^{\rm EW}$. For a full calculation of the pure one--loop QED
contributions we refer the reader
to the articles in Refs.~\cite{berends} and~\cite{denner}.

The analytical results for $\delta^{\rm EW}$ obtained in~\cite{bohm}
and~\cite{denner} involve huge formulae. We will show that within the
framework of our EFT, the analysis of the EW corrections
for these processes can be carried out with good accuracy in a very simple
way. Our approach just requires the knowledge of tree--level expressions
for the corresponding cross sections, plus the introduction of a few
input parameters, which can be fitted from existing experimental data for
LEP1 and SLD observables. The quality of the approach can be tested by
comparing our results with the full one--loop calculations for
$\delta^{\rm EW}$ mentioned above.

The paper is organized as follows: in section~\ref{sec:effective} we briefly
discuss how to integrate the top quark to obtain an effective Lagrangian
that can be used at LEP energies, and give explicit formulae for the effective 
couplings in the limit of large $\mt$. In section~\ref{sec:obseff} we use the 
previously obtained effective Lagrangian to calculate, in the large $\mt$
limit, the dominant radiative corrections to \ZZ and \gz. We show that this
simple tree--level analytical calculation in EFT is able to reproduce,
in this limit, the results obtained through a full one--loop calculation
in the SM.
In sections~\ref{sec:globalfit} and \ref{sec:lep2} we go beyond the large 
$\mt$ limit, assuming that our effective Lagrangian is valid for both LEP1 
and LEP2 energies. In section~\ref{sec:globalfit} we consider different 
$Z$--pole observables measured at 
LEP1 and SLD, and use tree--level
formulae (plus standard QCD and QED corrections) expressed in terms of the
effective Lagrangian couplings to get the corresponding theoretical
predictions. Then we use the experimental results 
of LEP1 and SLD to fit the parameters of the Lagrangian. Finally
the same Lagrangian is used in section~\ref{sec:lep2} to give predictions 
for \ZZ and \gZ at
LEP2 energies using again tree--level formulae. We expect 
to include in this way all the leading EW corrections to the observables
studied. This is checked by comparing our results with full one--loop
calculations. To conclude, in section~\ref{sec:conclusions} we collect the
main results of this paper.

\section{An effective EW Lagrangian for LEP energies}
\label{sec:effective}

The effective Lagrangian for $\m \le \mt$ is obtained by integrating
the top quark at $\mu=\mt$. At one--loop level this is done by
computing all diagrams containing at least one top quark. In the
case of the kinetic terms of the gauge bosons it is enough to compute
a few gauge boson self--energies. After a trivial field redefinition 
one obtains (for details see \cite{santi})
\begin{eqnarray}
{\cal L}_{\mathrm{eff}} &=& W_\mu^+ \partial^2 {W^-}^\mu+
{g_+^2(\m)\over 4} \left( v^2 +
\delta v_+^2(\mu)\right) W^+_\mu {W^-}^\mu +
{1\over 2} W^3_\mu \partial^2 {W^3}^\mu + {1\over 2} B_\mu \partial^2 B^\mu 
\nonumber \\
&+&{1\over 2} \left(g_3(\mu) W_3^\mu - g'(\mu)  B^\mu \right)
\left[{1\over 4}
\left(v^2+\delta v^2_3(\mu)\right)-\delta Z_{3Y}(\mu) \partial^2 \right]
\left(g_3(\mu) {W_3}_\mu - g'(\mu) B_\mu \right) 
\nonumber \\
&+&\ \bar \psi \ i
\Dslash(\ g_+(\mu)W^+,\ g_3(\mu)W_3,\ g'(\mu)B)\ \psi 
\nonumber \\
&+& i \bar{b}\dslash b +
\frac{1}{2}\left(g_3(\mu) W_3^\mu - g'(\mu)  B^\mu \right)(1+\epsilon_b(\mu))
\overline{b_L}\gamma_\mu b_L+\frac{1}{3} g'(\mu) B^\mu \bar{b}\gamma_\mu b~.
\label{eq:orglagr}
\end{eqnarray}
Here quark mixing has been neglected and $\psi$ stands for all the fermions
but the bottom and top quarks. Since the top quark has been integrated out,
there are no charged current couplings for the bottom. In addition,
the standard neutral couplings of the bottom quark get further modified due
to both vertex and wave function corrections to $b_L$. The contributions to
the $b_L$ self--energy have been absorbed
in (\ref{eq:orglagr}) by a redefinition of the $b_L$ field, whereas
the remaining corrections are collected in $\epsilon_b(\mu)$. The mixing
between the $W_3$ and $B$ wave functions has been
treated by including a $\partial^2$ operator in the form of a ``mass term"
to make simpler the subsequent diagonalization. The covariant derivative
$\Dslash$ in \eqn{eq:orglagr} is just a simplified notation to refer to the
standard gauge interactions to the fermions, but with the renormalized
couplings $g_+(\m)$, $g_3(\m)$ and $g'(\m)$. 

Higher dimensional operators suppressed by the corresponding inverse
powers of the top--quark mass, as well as other operators not relevant for
the discussion in this paper ---{\em e.g.} four fermion operators involving
the bottom quark--- have not been included. Triangle diagram contributions
to gauge boson interactions depending on the top--quark mass are small and
have also been neglected. In addition, in (\ref{eq:orglagr}) we have not
included trilinear couplings of gauge bosons; these will be important
for some processes such as $W$ boson production at LEP2.

The redefinition of fields also leads to a redefinition of
coupling constants. The initially unique coupling constant
$g$ splits into $g_+$ and $g_3$ below the top--quark mass
scale~\cite{peccei}. At one loop one obtains
$g_+^2(\mu) \simeq   g^2 \left(1-g^2 \delta Z_+(\mu)\right)$,
$g_3^2(\mu) \simeq  g^2\left(1-g^2 \delta Z_3(\mu)-g^2\delta
Z_{3Y}(\mu)\right)$ and
$g'^2(\mu) \simeq  g'^2\left( 1-g'^2 \delta Z_Y(\mu) -g'^2 \delta
Z_{3Y}(\mu)\right)$, where $g^2 \delta Z_+(\mu)$, $g^2 \delta Z_3(\mu)$, 
$g'^2 \delta Z_Y(\mu)$ are the top--quark induced wave function
renormalizations of the $W^+$, $W_3$ and $B$ gauge bosons respectively
while $g' g \delta Z_{3Y}(\mu)$ is the top--quark induced $W_3-B$
wave--function mixing.
Similarly $v_+^2(\mu)\equiv v^2 +\delta v^2_+(\mu)$ and 
$v^2_3(\mu)\equiv v^2+\delta v_3^2(\mu)$ are also different. 

In order to obtain the effective Lagrangian at LEP scales $\m\simeq  m_Z$
it is necessary to perform the matching of the effective theory
to the full theory at scales $\m=\mt$, and then to scale down
the effective Lagrangian, using the renormalization group equations,
for each of the ``couplings''
$g_+(\m),g_3(\m),g'(\m),\delta v_+^2(\m), \delta v_3^2(\m)$,
$\delta Z_{3Y}(\m)$ and $\epsilon_b(\m)$. In the limit of large
$\mt$ one obtains~\cite{santi}
\begin{mathletters}
\begin{eqnarray}
\frac{v^2_+(\mz)}{v^2_3(\mz)} & \simeq &
\frac{v^2_+(\mt)}{v^2_3(\mt)} \simeq
1+{3\over \4pi^2}{\mt^2\over v^2}
\label{eq:v+} \\
\frac{g^2_+(\mz)}{g^2_3(\mz)} & \simeq & 
1 + \frac{2g^2}{3\4pi^2} \logm
\label{eq:g+} \\
\delta Z_{3Y}(\mz) & \simeq & -\frac{1}{3\4pi^2} \logm
\label{eq:z3y} \\
\epsilon_b(\mz) & \simeq & 
-2\frac{\mt^2}{\4pi^2 v^2}-
\frac{1}{\4pi^2}\left(\frac{17}{6} g^2+\frac{1}{6} g'^2\right)\logm \;.
\label{eq:eb}
\end{eqnarray}
\label{eq:shifts}
\end{mathletters}

One observes here the leading non--decoupling top mass effects,
appearing both in the
universal self--energy coupling~\cite{veltman} and in the specific
vertex to $b$ quarks~\cite{akhundov}. QCD corrections to the parameters
in (\ref{eq:shifts}) can be easily included, if
needed~\cite{santi,santam} (for QCD corrections
to electroweak parameters in the large $m_t$ limit 
see also~\cite{qcdmt} and references therein).
Finally, to get the effective Lagrangian
at the $m_Z$ scale one still has to diagonalize
the neutral gauge boson sector, including a further wave function
renormalization of the $Z$ field to absorb the $\delta Z_{3Y}$ term
({\em i.e.} the mixing between the $W_3$ and $B$ wave functions).
The effective Lagrangian reads
\begin{eqnarray}
{\cal L}_{\mathrm{eff}} &=& W_\mu^+ \partial^2 {W^-}^\mu +
m_W^2 W^+_\mu {W^-}^\mu +
{1\over 2} A_\mu \partial^2 {A}^\mu + {1\over 2} Z_\mu \partial^2 Z^\mu 
+{1\over 2} \mz^2 Z_\mu {Z}^\mu \nonumber \\
&+&\ \bar \psi \ i
\Dslash(\ \frac{e_W(\mz)}{s_Z} W^+,\ \frac{e_Z(\mz)}{s_Z c_Z} Z,\ e(\mz)A)\ \psi \nonumber \\
&+& i \bar{b}\dslash b -
\frac{e_Z(\mz)}{2 s_Z c_Z}
\bar{b}\gamma_\mu \left(g^b_V-g^b_A \gamma_5\right) b Z^\mu
+\frac{1}{3} e(\mz)  \bar{b}\gamma_\mu b A^\mu~,
\label{eq:lagr}
\end{eqnarray}
where $c_Z\equiv \cos\theta_W(\mz)$, $s_Z\equiv \sin\theta_W(\mz)$,
are the cosine and the sine, respectively, of the effective weak mixing angle 
at the scale $\mz$. As usual, $\theta_W(m_Z)$ is determined by the
diagonalization of the mass matrix for the neutral gauge bosons. It
is trivially related to the gauge couplings by
$\tan\theta_W(\mz)\equiv g'(\mz)/g_3(\mz)$. In the same way,
$e(\mz)= g_3(\mz) s_Z$
is the electromagnetic coupling at the scale $\mz$. As commented above, the
$Z$ field needs a further rescaling owing to the $\delta Z_{3Y}$ term
in (\ref{eq:orglagr}). This leads us to define $e_Z(\mz)$, which
appears in all Z couplings and which is related to $e(\mz)$ through
\bq
e^2_Z(m_Z)= e^2(m_Z) \left(1-\frac{g^2}{c_Z^2} \delta Z_{3Y}(\mz)\right)
\label{eq:ez}~.
\eq
Similarly, we found it convenient to express the coupling of the
$W^+$ gauge bosons, $g_+(\mz)$, in terms of an effective coupling
$e_W(\mz) \equiv g_+(\mz) s_Z$. From this definition and Eq.\
(\ref{eq:g+}) we get
\bq
e^2_W(m_Z)= e^2(m_Z) \frac{g^2_+(\mz)}{g^2_3(\mz)}\;.
\label{eq:ew}
\eq
As it is usually done for the electromagnetic coupling, we can define
\begin{eqnarray}
\alpha(\mz) \equiv  \frac{e^2(\mz)}{4\pi} \equiv  \frac{\alpha}{1-\Delta\alpha}
\ , \label{eq:alpha}\\
\alpha_Z(\mz) \equiv  \frac{e^2_Z(\mz)}{4\pi} \equiv  
\alpha(\mz) \left(1+\delta\alpha_Z\right)
\ ,\label{eq:alphaZ}\\
\alpha_W(\mz) \equiv \frac{e^2_W(\mz)}{4\pi} \equiv  
\alpha(\mz) \left(1+\delta\alpha_W\right)~.
\label{eq:alphaW}
\end{eqnarray}
Here $\alpha=1/137.036$ is the fine structure constant and
$\Delta \alpha$ is the QED shift produced by the running
from its on--shell value to $\m=\mz$. It can be obtained from
$e^+e^- \rightarrow {\em hadrons}$ data. If $\alpha(\mz)$ is given in
the \msb scheme one obtains $\Delta\alpha=0.067$~\cite{kniehl,PDG}.
$\delta\alpha_Z$ and $\delta\alpha_W$ represent the additional
shifts in $\alpha_Z$ and $\alpha_W$ due, in part, to the heavy top. From
(\ref{eq:g+}), (\ref{eq:z3y}), (\ref{eq:ez}) and (\ref{eq:ew}) we
obtain
\begin{eqnarray}
\delta\alpha_Z &\simeq&  
\frac{\alpha}{12\pi s^2_Z c^2_Z}\logm~,  \label{eq:dalphaZ}\\ 
\delta\alpha_W &\simeq&
\frac{\alpha}{12\pi s^2_Z}\logm~.   \label{eq:dalphaW}     
\end{eqnarray}
Given the size of these corrections, we expect the three couplings
$\alpha(\mz)$, $\alpha_Z(\mz)$, and $\alpha_W(\mz)$ to be almost equal, 
at least, at the percent level. Their values can be extracted directly from 
experiment and we will see that, indeed, that is the case. Therefore, if only
precisions at the percent level are needed, one can safely assume
$\alpha(\mz)=\alpha_Z(\mz)=\alpha_W(\mz)$.

The physical $W$ and $Z$ masses are given, in the large $\mt$ limit, by the 
equations
\begin{eqnarray}
m_W^2 &=& \frac{e^2_W(\mz)}{4 s^2_Z} \ v^2_+(\mz) \nonumber \\
m_Z^2 &=&  \frac{e_Z^2(\mz)}{4 c^2_Z s^2_Z} v^2_3(\mz)~.
\label{eq:masses}
\end{eqnarray}
Then, we can obtain the cosine of the Sirlin weak mixing angle~\cite{sirlin} 
in terms of the cosine of the effective mixing angle at the scale $\mz$:
\bq
c^2_W \equiv \frac{m^2_W}{m^2_Z}=
c^2_Z \frac{e^2_W(\mz)}{e^2_Z(\mz)}\frac{v^2_+(\mz)}{v^2_3(\mz)}~.
\label{eq:cw}
\eq
If we write now the relation between these two mixing angles as
$s^2_Z=s^2_W+\delta s^2_W$, from equations~(\ref{eq:v+}--\ref{eq:z3y}),
(\ref{eq:ez}) and  (\ref{eq:ew}) we immediately obtain
\bq
\delta s^2_W = \frac{\alpha}{\pi}\left[\frac{3}{16 s^2_Z}
\frac{m_t^2}{m_Z^2}+\frac{(3-2 s^2_Z)}{12 s^2_Z}
\log\left(\frac{m_t}{m_Z}\right)\right]~.
\label{eq:dsw}
\eq
On the other hand, the coupling of the bottom quark to the $Z$ boson gets 
extra contributions due to the vertex diagrams involving the top quark.
These contributions can be taken into account by parametrizing the
effective $g^b_V$ and $g^b_A$ couplings as
\bq
g^b_V = -\frac{1}{2}\left(1+\epsilon_b(\mz)\right) +\frac{2}{3} s^2_Z\ , \ \ \ \ 
g^b_A = -\frac{1}{2}\left(1+\epsilon_b(\mz)\right)  \ , \ \ \ \ 
\label{eq:gbs}
\eq
where $\epsilon_b(\mz)$ is given in \eqn{eq:eb}.
Finally, although it is not relevant for the neutral current processes
we want to study, one can also relate $m_W$ and $s_Z$ with the
Fermi coupling constant, $\gf$, measured in the muon decay~\cite{stuart}. 
This relation reads
\bq
{\gf \over \sqrt2} \simeq {e^2_W(\mz)\over 8 m_W^2 s^2_Z}~,
\label{eq:gf}
\eq
and allows to estimate $\alpha_W(m_Z)$ from $s_Z$, $m_W$ and $G_F$.

\section{\ZZ and \gZ in the large $\mt$ limit}
\label{sec:obseff}

We can use now the Lagrangian (\ref{eq:lagr}) at tree level, together with
the results in the previous section, to estimate the dominant electroweak
corrections to \ZZ and \gZ at LEP2 in the large $\mt$ limit.

\begin{figure}[htbp]
\begin{center}
\vspace{1cm}
\epsfig{file=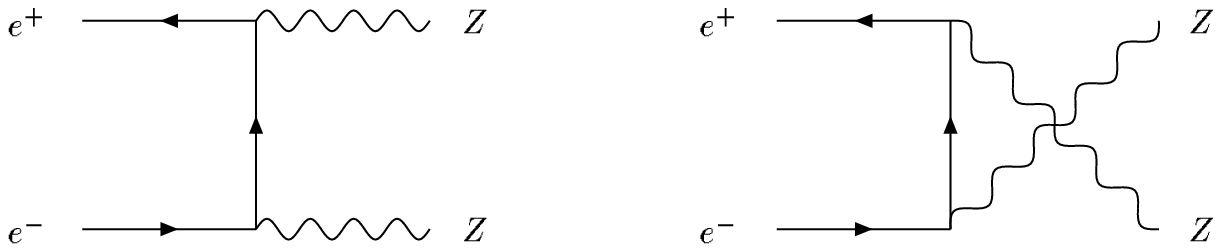}
\end{center}
\vspace{1cm}
\caption{Tree--level contributions to $e^+ e^-\rightarrow Z Z$.
\label{fig:zz}}
\end{figure}

{}From the Lagrangian (\ref{eq:lagr}) and the diagrams in
Fig.~\ref{fig:zz}
we easily obtain the cross section for \zz. It is the usual tree--level
result obtained in the SM, but expressed in terms of the
effective couplings $\alpha_Z(\mz)$ and $s_Z$:
\begin{eqnarray}
\left(\frac{d\sigma^{ZZ}}{d\Omega}\right)_{\mathrm{eff}} & = &
\frac{\alpha^2_Z(\mz)}{32 s^4_Z c^4_Z}
\,(g^{4}_V+6 g^{2}_V g^{2}_A+g^{4}_A)
\nonumber \\
& & \times \,
\frac{1}{s} \left(1-\frac{4m_Z^2}{s}\right)^\frac{1}{2}
\left[\frac{s^2+6 m_Z^4}{u t}-\frac{m_Z^4(s-2 m_Z^2)^2}{(u t)^2}-2\right]\,,
\label{eq:zz}
\end{eqnarray}
where $g_V=-1/2+2 s^2_Z$, $g_A=-1/2$ are, respectively, the vector and 
the axial--vector couplings of the $Z$ boson to the electrons,
and $s$, $t$ and $u$ are the usual Mandelstam variables
\bq
s=(p_++p_-)^2\,,\quad\quad t=(p_+-p_1)^2 \,,\quad\quad
u=(p_+-p_2)^2\,,
\label{eq:stu}
\eq
being $p_+$, $p_-$ and $p_1$, $p_2$ the lepton and gauge boson four--momenta
respectively. In the center of mass frame, the dependence on the scattering
angle $\theta$ is carried by $t$ and $u$ through the relations
\begin{eqnarray}
t & = & -\frac{1}{2}\left[s-m_1^2-m_2^2-\lambda^{1/2}(s,m_1^2,m_2^2)
\cos\theta\right] \;, \nonumber\\
u & = & -\frac{1}{2}\left[s-m_1^2-m_2^2+\lambda^{1/2}(s,m_1^2,m_2^2)
\cos\theta\right]\;,
\label{eq:tu}
\end{eqnarray}
where $\lambda(s,m_1^2,m_2^2)\equiv (s-m_1^2-m_2^2)^2-4m_1^2m_2^2$
(here $m_1=m_2=m_Z$).

The accuracy of the effective cross section (\ref{eq:zz})
can be tested by comparing with full explicit
EW calculations in the SM. As commented in the introduction, 
the EW corrections to \ZZ were calculated in~\cite{denner}. There,
the Born cross section is defined in terms of the fine structure constant 
$\alpha$ and the Sirlin weak mixing angle $s_W$. That is, our expression
(\ref{eq:zz}), but changing $\alpha_Z(\mz) \rightarrow \alpha$ and
$s_Z \rightarrow s_W$. With respect to this reference value, the effective
cross section (\ref{eq:zz}) shows a correction
\bq
\delta^{\rm{EW}\,(ZZ)}_{\rm eff}\equiv
\frac{\left(\frac{d\sigma^{ZZ}}{d\Omega}\right)_{\rm eff}
-\left(\frac{d\sigma^{ZZ}}{d\Omega}\right)_0}
{\left(\frac{d\sigma^{ZZ}}{d\Omega}\right)_0} \;.
\label{eq:dzz}
\eq
To compare the results from our effective Lagrangian with those
in~\cite{denner} we rewrite \eqn{eq:zz} in terms of $\alpha$ and $s_W$
by using (\ref{eq:alphaZ}) and $s^2_Z = s^2_W+\delta s^2_W$.
The leading electroweak corrections can be easily obtained in the
large $\mt$ limit:
\bq
\delta^{\rm{EW}\,(ZZ)} \simeq 
2\,\Delta\alpha+
2\,\delta\alpha_Z+
\frac{s^4_Z c^4_Z}{(g_V^4+6g_V^2 g_A^2+g_A^4)}
\,\frac{d\mbox{ }}{d s^2_Z}
\left[\frac{(g_V^4+6g_V^2 g_A^2+g_A^4)}{s^4_Z c^4_Z}\right]
\delta s^2_W \,,
\label{eq:dzz1}
\eq
with $\delta\alpha_Z$ and $\delta s^2_W$
given by (\ref{eq:dalphaZ}) and (\ref{eq:dsw}) respectively. 
Using these equations we find
\begin{eqnarray}
\delta^{\rm{EW}\,(ZZ)} & \simeq & 2\,\Delta\alpha
-\frac{\alpha}{\pi s_Z^4 c_Z^2}\left[
{\frac{3\,\left(1 - 6\,{s_Z^2} + 12\,{s_Z^4} - 8\,{s_Z^6} -16\,{s_Z^8} \right)}
 {8\,\left( 1 - 8\,{s_Z^2} + 24\,{s_Z^4} - 32\,{s_Z^6} + 32\,{s_Z^8} \right) }}
  \,\frac{\mt^2}{\mz^2}+\right. \nonumber \\
& & + \left.
{\frac{\left(3 - 21\,{s_Z^2} + 56\,{s_Z^4} - 72\,{s_Z^6} \right)}
 {6\,\left( 1 - 8\,{s_Z^2} + 24\,{s_Z^4} - 32\,{s_Z^6} + 32\,{s_Z^8} \right) }}
  \logm
\right]\,.
\label{eq:dzz2}
\end{eqnarray}
We have checked explicitly this result against the one obtained
in~\cite{denner} by taking there the large $\mt$ limit,
and found complete agreement.

\begin{figure}[htbp]
\begin{center}
\vspace{1cm}
\epsfig{file=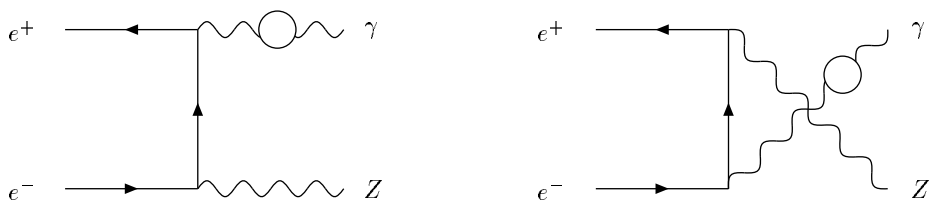}
\end{center}
\vspace{1cm}
\caption{Photon self--energy diagrams contributing to $e^+ e^-\rightarrow
\gamma Z$.
\label{fig:gz}}
\end{figure}

Let us discuss now the case of $\gamma Z$ production at LEP2. 
As in the previous case, we begin by writing the SM lowest--order
differential cross section for the process, which is obtained from
the diagrams in Fig.~\ref{fig:zz} after replacing one of the $Z$ bosons by a 
photon. Here we find
\bq
\left(\frac{d\sigma^{\gamma Z}}{d\Omega}\right)_{\rm eff} = 
\frac{\alpha \alpha_Z(\mz)}{4 s^2_Z c^2_Z}\,
(g^{2}_V+g^{2}_A)\frac{(s-m_Z^2)}{s^2}
\left(\frac{s^2+m_Z^4}{2 u t}-1\right)\,.
\label{eq:gz}
\eq
Although in principle the situation is similar to the case of
$ZZ$ production, there is a crucial difference if the photon is
a real one, that is with $q^2=0$. In that
case, in the \msb scheme we are using, and choosing a
renormalization scale $\m=\mz$, one finds that the photon self--energy
diagrams of Fig.~\ref{fig:gz} contain large logarithms
$\sum_f Q^2_f \log(m_f/\mz)$, which effectively produce the
``running back'' of the electromagnetic coupling constant from
$\alpha(\mz)$ to $\alpha(m_e) \simeq \alpha$. This agrees with
common wisdom that says that real, on--shell photons (no matter whether
they are soft or hard) couple with
$\alpha$ strength. Thus, the use of our effective Lagrangian
at the scale $\m=\mz$ has to be supplemented with the rule that 
a real photon couples with its on--shell coupling.

As in the previous case, we can easily obtain the electroweak correction
\bq
\delta^{\rm{EW}\,(\gamma Z)} = \Delta\alpha+\delta\alpha_Z+
\frac{s_Z^2 c_Z^2}{(g_V^2+g_A^2)}\,\frac{d\mbox{ }}{d s_Z^2}
\left[\frac{(g_V^2+g_A^2)}{s^2_Z c^2_Z}\right]
\delta s^2_W \,,
\label{eq:dgz1}
\eq
and using (\ref{eq:dalphaZ}) and (\ref{eq:dsw}) we get, in the large
$\mt$ limit,
\begin{equation}
\delta^{\rm{EW}\,(\gamma Z)}= \Delta\alpha
-\frac{\alpha}{\pi s_Z^4 c_Z^2}\left[
{\frac{3\left(1 - 2\,{s_Z^2} - 4\,{s_Z^4} \right) }
    {16\,\left( 1 - 4\,{s_Z^2} + 8\,{s_Z^4} \right) }}
    \frac{\mt^2}{\mz^2}+  
    {\frac{\left(3 - 9\,{s_Z^2} - 4\,{s_Z^4} \right) }
    {12\,\left( 1 - 4\,{s_Z^2} + 8\,{s_Z^4} \right) }}
    \logm
\right]\,.
\label{eq:dgz2}
\end{equation}

As in the $ZZ$ production, we should be able to test our EFT approach
by comparing our results with those arising from explicit
calculations of the EW effects in the limit of large $\mt$.
For this process, an explicit one--loop calculation of
$\delta^{\rm{EW}\,(\gamma Z)}$ in the SM has been carried out
by B\"ohm and Sack~\cite{bohm}.
However, we could not reproduce their results in the large $\mt$ limit.
We believe they missed a factor $1/2$ entering the $Z$ boson renormalized
self--energy, therefore the bulk of the EW corrections, namely
$\Delta \alpha$, has been overestimated. This is
confirmed by the study of the crossed reaction
$e^-\gamma\rightarrow e^- Z$, which has been extensively analysed
in the literature (see for instance \cite{dittmaier}). Here the EW
corrections are reproduced completely, in the large $\mt$ limit,
with \eqn{eq:dgz2}\footnote{Notice that in our approach the size of the
electroweak corrections is exactly the same in the case of \gZ and
$e^-\gamma\rightarrow e^- Z$.}.

\section{Global fit for LEP1/SLD observables}
\label{sec:globalfit}
The analytical results of the previous section, obtained in the large
$\mt$ limit, are very interesting and very useful to test the overall
approach and contain the bulk of EW radiative corrections. However,
the top--quark mass is not so large and there could be other corrections
at least comparable to those considered. In addition, Higgs mass
corrections, though in principle can also be included, have not been
taken into account in the previous analysis. This
makes it difficult to achieve precisions better than the 2--3\% with the
above analytical approach. As an alternative procedure, we can use our
effective Lagrangian at tree level with arbitrary couplings, and fit those 
couplings with LEP1/SLD observables. In this way, the effective couplings will
contain not only the leading top--quark and Higgs mass dependences but also
other universal non--leading corrections. This includes, a priori,
also possible effects of new physics in the effective couplings. In this
sense, the procedure is related to the ``$S,T,U$''~\cite{peskin} or
``epsilon''~\cite{altarelli} analyses proposed already in the literature.
The excellent agreement between the SM predictions and LEP1/SLD observables
suggests, however, that our EFT couplings do not include any
significant effect arising from non--standard physics.

Once we have fitted the parameters entering the effective Lagrangian for
the processes observed at LEP1/SLD, we can use the result to give predictions
for \ZZ and \gZ cross sections at LEP2 energies, considering once again
tree--level formulae.
With this procedure, we expect to achieve precisions better than
1\%, which should be enough for most LEP2 observables. This can be checked
by comparing our results with known one--loop calculations.

The list of LEP1 and SLD observables that we consider for our
fit is presented in Table I. These include: the $Z$ mass
($m_Z$), the total $Z$ width ($\Gamma_Z$), the hadronic cross section
($\sigma_{\rm had}$), the ratio of the widths $Z\to {\em hadrons}$ to
$Z\to l^+l^-$, $l=e,\mu,\tau$ ($R_l$), the ratios of the widths $Z\to \bar b b$
($R_b$) and $Z\to \bar c c$ ($R_c$) to $Z\to{\em hadrons}$, the leptonic
($A_{FB}^{(0,l)}$),
$b$--quark ($A_{FB}^{(0,b)}$) and $c$--quark ($A_{FB}^{(0,c)}$) $C$--odd
forward--backward asymmetries, and the $P$--odd leptonic ($A_l$), b--quark
($A_b$) and c--quark ($A_c$) asymmetries. The quoted experimental value
for the leptonic asymmetry $A_l$ is the average of LEP1 and SLD results,
assuming lepton universality.  

{}From the Lagrangian (\ref{eq:lagr}), it is immediate
to see that the lowest--order formulae in the EFT scheme are basically the same
as in the SM, just taking $e_Z(m_Z)/(s_Z c_Z)$ and $s_Z$ as the weak 
$Z\bar ff$ coupling constant and the sine of the Weinberg angle respectively.
Only special care has to be taken in the case of the $Z\bar bb$ coupling, which
requires the inclusion of the additional parameter $\epsilon_b(m_Z)$ defined 
in the previous section. The tree--level expressions for the LEP1/SLD
observables, as well as the leading QCD and QED corrections, are well--known
and will not be reproduced here.

The parameters to be fitted are five, namely $m_Z$, $\alpha_Z(m_Z)$,
$s^2_Z$, $\epsilon_b(m_Z)$ and $\alpha_s(m_Z)$, although the value of
$\alpha_s(\mz)$ could be obtained independently from other
processes~\cite{PDG}.
In any case, the result of the fit is found to be rather stable with respect
to the value of $\alpha_s(\mz)$.

We obtain from the fit the following values for the parameters:
\begin{eqnarray}
m_Z & = & 91.1867\pm 0.0020  \nonumber\\
\alpha_Z(m_Z) & = & 0.007788 \pm 0.000012 \nonumber\\
s^2_Z & = & 0.23103\pm 0.00021  \nonumber\\
\epsilon_b(m_Z) & = & -0.0053\pm 0.0023 \nonumber\\
\alpha_s(\mz) & = & 0.1215 \pm 0.0052
\label{eq:fit}
\end{eqnarray}
with
\bq
\chi^2/\rm{ndf} = 2.6/7\,.
\label{eq:chi2}
\eq
If instead we fix $\alpha_s(\mz)$ to its world average 
$\alpha_s(\mz)=0.119$~\cite{PDG} we obtain 
$m_Z=91.1867\pm 0.0020$, $\alpha_Z(m_Z)=0.007790 \pm 0.000011$,
$s^2_Z=0.23102\pm 0.00021$, and
$\epsilon_b(m_Z)=-0.0045\pm 0.0017$ with
$\chi^2/\rm{ndf}=2.9/8$.

Notice that in these fits we neglect the correlations
in the input data. We have checked that the impact of these
correlations in the results of the fit is negligible. We did so by 
performing a fit including the main correlations (we have 
considered only the correlations which are larger than 10\%~\cite{ewwg99}: 
$\Gamma_Z$--$\sigma_{\rm had} \approx -0.19$,
$R_l$--$\sigma_{\rm had} \approx 0.13$,$R_b$--$R_c \approx -0.17$,
$A_{FB}^{(0,b)}$--$A_{FB}^{(0,c)} \approx 0.13$). The errors remain unchanged
and the central values are shifted at most by 10\% of one standard 
deviation.

The predictions for the LEP1/SLD observables obtained with these values can be
read from the third column in Table I. We also quote in the fourth column
the deviations of the different observables from the measured central
values in units of experimental standard deviations (the pull). It can be 
seen that all the predictions deviate less than 
1.5$\sigma$ from the measured values. This is reflected in the very low
$\chi^2$ in (\ref{eq:chi2}), and shows that for LEP1 and SLD data the EFT
approach works remarkably well. 

{}From \eqn{eq:gf}, using the value of $s^2_Z$ in (\ref{eq:fit}), the $W$
boson mass, $m_W=80.42\pm 0.08$~GeV, and the Fermi constant, we can
also estimate the value of $\alpha_W(\mz)$. We obtain
$1/\alpha_W(\mz)=127.2\pm 0.3$, to be compared with
$1/\alpha_Z(\mz)=128.4\pm 0.2$ from our fit (\ref{eq:fit}), and to
$1/\alpha(\mz)=127.88\pm 0.09$ obtained by running from the Thomson
limit. We see that, as expected from (\ref{eq:dalphaZ}) and
(\ref{eq:dalphaW}), the differences are really small and the three couplings
can be taken as equal if only precisions at the 1\% level are needed. 

\section{Predictions for \ZZ and \gZ at LEP2}
\label{sec:lep2}

Now, once the effective couplings at the
scale of $m_Z$ have been determined, our goal is to use the same approach
to predict the magnitude of the electroweak corrections for processes to be
measured at LEP2. To estimate the corresponding cross sections, we will
use the values of $\alpha_Z(m_Z)$, $m_Z$ and $s^2_Z$ from the result of the
fit (\ref{eq:fit}). Although the relevant scale at LEP2 could reach
190~GeV, the running of the parameters from $\mz$ to 190~GeV
will give at most corrections of the order of $\alpha/\pi\log 2$
which are small\footnote{Note, however, that if needed, these corrections
can be easily included in our approach.}. In addition, in the processes
we are interested in this paper, \ZZ and \gz, the gauge bosons are
on--shell, therefore the relevant scale is fixed by their masses.

Let us take the EFT tree--level expressions (\ref{eq:zz}) and (\ref{eq:gz})
for \ZZ and \gZ respectively, with $\alpha_Z$, $m_Z$ and $s^2_Z$
from (\ref{eq:fit}), and compute the size of the deviations from
the Born cross sections expressed in the on--shell scheme as in
\eqn{eq:dzz} (or its equivalent for $\gamma Z$ production). 
For the process \ZZ we obtain
\bq
\delta^{\rm{EW}\,(ZZ)}_{\rm eff}
\simeq 5.4\pm 0.4\ \%\,.
\label{eq:dzzfit}
\eq
Since for our fit we have used LEP1/SLD values, we have also taken the
last SM fit for $Z$--pole data to evaluate the Born cross section
written in terms of on--shell parameters. Thus, we have used for
the Sirlin weak mixing angle the value $s^2_W = 0.2236\pm 0.0008$~\cite{PDG}.

On the other hand, the full one--loop EW correction to $(d\sigma/d\Omega)_0$
can be obtained from the analysis carried out by Denner and Sack~\cite{denner},
after updating the
values for the masses of the gauge bosons and ---especially--- the top quark.
It can be seen that, for LEP2 energies, the shift is strongly dominated
by the one--loop correction to the $Z$ self--energies, which contain the top
quark dependence. To obtain the EW corrections from \cite{denner} we use
the value of $s^2_W$ quoted above, together with
$m_t=168\pm 8$~GeV (arising from $Z$--pole analysis~\cite{PDG})
and $m_H=m_Z$. We find
\bq
\delta^{\rm{EW}\,(ZZ)}_{\rm{SM-1\ loop}}
\simeq 5.3\pm 1.0\ \%\, ,
\label{eq:dzz1loop}
\eq
where the error is mainly due to the uncertainty in $m_t$. As can be seen,
the agreement between the values in (\ref{eq:dzzfit}) and (\ref{eq:dzz1loop})
is remarkably good. Notice that, in general,
one would expect $\delta^{\rm{EW}\,(ZZ)}_{\rm{SM-1\ loop}}$ to depend on the
scattering angle $\theta$. However, it can be seen~\cite{denner} that for
these energies the distribution is almost flat, so that the constant value in
(\ref{eq:dzz1loop}) represents a good approximation. This is also consistent
with our approach. As can be seen from Eqs.\ (\ref{eq:zz}) and (\ref{eq:tu}),
the dependence of the differential Born cross section with the scattering
angle is contained in the Mandelstam variables $t$ and $u$, which only
enter the factor in square brackets in (\ref{eq:zz}). The shift of $\alpha$
and $\sin\theta_W$ from the on--shell to the effective values leads only to a
global correction that does not affect the $\theta$--dependence.

It is important to remark that the value in (\ref{eq:dzzfit}) has been found
in a quite straightforward way, whereas that in (\ref{eq:dzz1loop}) can be
obtained only after a very lengthy calculation. In addition, the one--loop
result, though in principle more precise, depends not only on $m_t$ but
also on other uncertain parameters, such as the Higgs mass and the running of
the electromagnetic coupling from the Thomson limit to the $m_Z$ scale.

\hfill

For \gZ we use \eqn{eq:gz}, taking once again $\alpha_Z$, $m_Z$ and
$s^2_Z$ from (\ref{eq:fit}). Notice that, as
discussed in the previous section, the value of the 
electromagnetic coupling to be used in (\ref{eq:gz}) is the on--shell fine
structure constant $\alpha$. This is because the photon is on shell,
{\em i.e.}\ with $q^2=0$. For the rest of the parameters the same
considerations as for \ZZ apply. We obtain
\bq
\delta^{\rm{EW}\,(\gamma Z)}_{\rm eff}
\simeq 3.7\pm 0.2\ \%\,.
\label{eq:dgzfit}
\eq
As commented in section~\ref{sec:obseff} we cannot use the expressions
in~\cite{bohm} to check this last result. Still, we can take into account
the known calculations~\cite{dittmaier} for the crossed reaction  
$e^-\gamma\rightarrow e^- Z$. Within the EFT approach, the cross section
for this process shows exactly the same dependence on the parameters
$\alpha_Z(\mz)$ and $s_Z$ as in (\ref{eq:gz}), therefore for both
$e^-\gamma\rightarrow e^- Z$ and \gZ the correction
$\delta^{\rm EW}_{\rm eff}$ will be exactly the same.

For a center--of--mass energy of 100 GeV, and using a top--quark mass
of 140 GeV, the
analysis in Ref.~\cite{dittmaier} shows that the EW corrections to
$e^-\gamma\rightarrow e^- Z$ are $\simeq 4.2$\%. Once again, it is found 
that this result is almost independent from the
scattering angle (see Table 3 of~\cite{dittmaier}). Now increasing $m_t$
up to 168~GeV, and taking into account the errors in $s_W$ and
$m_t$ as in the \ZZ case, we find 
\bq
\delta^{{\rm EW}\,e^-\gamma\rightarrow e^- Z}_{\rm SM-1\ loop}
\simeq 3.1\pm 0.4\%\,.
\label{eq:dez1loop}
\eq
That means, our result (\ref{eq:dgzfit}) lies within the expected level of
accuracy. Moreover, our approach succeeds
in predicting the flat behaviour of $\delta^{\rm EW}$ with respect to
the scattering angle.

Notice that in the definition of $\delta^{\rm EW}$ we
refer to the Born cross section. The latter is defined in terms
of the Sirlin weak mixing angle (or equivalently, the $W$ mass), which
is not measured with sufficiently high precision in LEP1/SLD and
introduces some error. In order to estimate the accuracy of
our approach, it is better to compare directly the values for the cross
sections obtained from both the EFT and SM one--loop analyses. In this way
the comparison is much less sensitive to the top--quark mass, which
does not appear explicitly in $\sigma_{\rm eff}$. Thus for both
\ZZ and $e^-\gamma\rightarrow e^- Z$ we compute the ratio
\bq
\Delta\equiv
\frac{\sigma_{\rm eff}-\sigma_{\rm SM-1\ loop}}{\sigma_{\rm SM-1\ loop}}\;,
\eq
obtaining
\bq
\Delta^{(ZZ)}=0.0012\pm 0.0038
\eq
and
\bq
\Delta^{(\gamma Z)}=0.0065\pm 0.0017\;.
\eq
In both cases, the agreement between EFT and one--loop SM values is found
to be better than 1\%. For the crossed reaction \gz, the result is expected
to be similar to that obtained for $e^-\gamma\rightarrow e^- Z$.

It is also worth to mention that in our effective Lagrangian approach
we trivially find that pure Compton processes (only containing real photons
and electrons), such as $e^-\gamma \rightarrow e^-\gamma$ and
$\gamma\gamma \rightarrow e^-e^+$, have zero EW radiative corrections, since
the tree--level cross sections are independent of $s_Z$ and $\alpha_Z(\mz)$.
It can be seen that this result is also obtained from full one--loop SM
calculations~\cite{compton}, for a range of center of mass energies
of 100--200~GeV. Our approach is again successful in this case.

\hfill

\section{Conclusions}
\label{sec:conclusions}

In this paper we elaborate an effective field theory approach to the
analysis of the electroweak corrections at LEP energies. 

We review how to obtain the effective EW Lagrangian that arises when the
top quark is integrated out. At the leading order in the top--quark mass,
we obtain the effective couplings that are relevant for LEP energies.

Using this Lagrangian at tree level we obtain analytical formulae for the
differential cross sections for the LEP2 processes \ZZ and \gz, in the large 
$m_t$ limit. The results agree completely with full one--loop EW
calculations.

This approach allows us to compute LEP2 observables in the large $m_t$
limit. However, this cannot be used in general to achieve precisions better
than 2--3\%. To go beyond that, we consider an effective Lagrangian similar
to that arising from the EFT, but leaving the couplings as free
parameters. Then, using the effective Lagrangian at tree level, we fit
the parameters from present LEP1 and SLD data. In this way, the effective
couplings
should take into account the effects arising from virtual top quarks
and Higgs, as well as other possible universal contributions. The fit
is performed for 12 LEP1/SLD observables, and the parameters to be determined
are five,
including $m_Z$ and $\alpha_s(m_Z)$. The results are amazingly good: as
shown in Table~\ref{tab:fit}, in all cases the difference between fitted
and experimental values is less than 1.5$\sigma$. Finally,
taking the effective couplings from the LEP1/SLD fit,
we compute the differential cross sections for \ZZ and \gZ at LEP2 energies,
using once again the effective Lagrangian at tree level. The predictions are
in this way completely independent on the masses of the top quark and the
Higgs boson. Our results are compared with the values of the corresponding
Born cross sections written in terms of
on--shell parameters: at LEP2 energies, the EW corrections for \ZZ and \gZ
amount to 5.4~\% and 3.7~\% respectively for a fixed value of the on--shell
weak mixing angle. In addition, our EFT cross sections are compared with
those arising from one--loop analyses in the SM. The agreement is found to
be better than 1~\% for both \ZZ and \gz.

It is worth to point out that our effective Lagrangian can be used to
estimate the
size of EW corrections in other LEP2 processes that involve the
subprocesses $e^+ e^- \rightarrow V V$, $V e \rightarrow V e$ and
$V V\rightarrow \bar{f} f$ ($V= \gamma, Z$). This is {\em e.g.} the
case of the scattering $e^+ e^- \rightarrow e^+ e^- \bar{b} b$, which
is a very important background process in searches for new particles,
and where the full one--loop EW calculation turns out to be very
hard . Another important example is the neutrino counting process 
$e^+ e^- \rightarrow \nu\bar{\nu} \gamma$,
for which the full one--loop EW calculation is also missing. The
extension of the effective Lagrangian to include triple gauge boson
couplings is presently under study.

\acknowledgements

A.\ A.\ A.\ thanks the Spanish Ministry of Education and Culture 
for the sabbatical grant No. SAB1998-0068 at the University of Valencia.
D.\ G.\ D.\ has been supported by a grant from the Commission
of the European Communities, under the TMR programme (Contract
No.\ ERBFMBICT961548). 
This work has been funded by CICYT under the Grant AEN-96-1718, 
by DGESIC under the Grant PB97-1261 and by the DGEUI of 
the Generalitat Valenciana under the Grant GV98-01-80.

\begin{table}
\caption{Results for combined LEP1 and SLD observables obtained within the EFT
scheme using the fitted values for $\alpha_Z(m_Z)$, $m_Z$, $s_Z$,
$\epsilon_b(m_Z)$ and $\alpha_s(m_Z)$. We have used EFT tree--level formulae
plus standard QCD and QED corrections.
\label{tab:fit}}
\begin{tabular}{cccc}
Observable & Experimental value & Fitted value & Pull
\\ \hline
$m_Z \mbox{ [GeV]}$ & $91.1867\pm 0.0020$ & 91.1867 & 0.00\\
$\Gamma_Z \mbox{ [GeV]}$ & $2.4948\pm 0.0025$ & 2.4949 & 0.04\\
$\sigma_{\rm had} \mbox{ [nb]}$ & $41.486\pm 0.053$ & 41.499 &  0.24 \\
$R_l$ & $20.775\pm 0.027$ & 20.779 & 0.15 \\
$R_b$ & $0.2170\pm 0.0009$ & 0.2169 & -0.09\\
$R_c$ & $0.1734\pm 0.0048$ & 0.1711 & -0.47\\
$A_{FB}^{(0,l)}$ & $0.0171\pm 0.0010$ & 0.0170 & -0.05 \\
$A_{FB}^{(0,b)}$ & $0.0984\pm 0.0024$ & 0.1015 & 1.30\\
$A_{FB}^{(0,c)}$ & $0.0741\pm 0.0048$ & 0.0726 & -0.31\\
$A_l$ & $0.1521\pm 0.0021$ & 0.1506 &  -0.71\\
$A_b$ & $0.900\pm 0.050$ & 0.899 & -0.02 \\
$A_c$ & $0.650\pm 0.058$ & 0.643 & -0.12\\
\end{tabular}
\end{table}

\end{document}